\documentclass[aps,prb,twocolumn,showpacs]{revtex4}
\usepackage{graphicx}
\begin{document}

\title{Vanishing Spin-Hall Conductivity in 2D disordered Rashba electron gas}

\author{Ol'ga V. Dimitrova}

\affiliation{L. D. Landau Institute for Theoretical Physics, Russian Academy 
of Sciences, Kosygina str. 2,  Moscow 119334, Russia}

\date{\today}

\begin{abstract}
For the two-dimensional ideal electron gas with the Rashba spin-orbit 
interaction and in the presence of non-magnetic short-ranged potential 
impurities the spin-Hall conductivity $\sigma_{sH}$ is found by direct 
microscopic calculation. Within the semiclassical approximation 
$\hbar/\tau \ll \epsilon_F$ and $\Delta \ll \epsilon_F$ 
the value of $\sigma_{sH}$ {\it is zero} for arbitrary ratio of the 
spin-orbit splitting $\Delta$ and the inverse elastic scattering time $1/\tau$.
\end{abstract}


\maketitle

Recently it has been proposed~\cite{Murakami} that a {\it dissipationless } 
spin current can be generated in response to an electric field in 
semiconductors with the spin-orbital interaction. For the case of an ideal 
two-dimensional (2D) electron gas with the Rashba coupling, 
Sinova et al.~\cite{Sinova} have found a spin-Hall current of the transverse 
($z$) spin component as a response to an in-plane electric 
field $E_{\nu}$, $j^z_\mu = \sigma_{sH} \epsilon_{\mu\nu} E_\nu$, with the 
``universal'' (independent upon the spin-orbital band splitting $\Delta$) 
spin-Hall conductivity $\sigma_{sH}= e/8\pi\hbar$. 
In the presence of disorder and for the simplest standard model it was 
demonstrated a cancellation of the spin-Hall effect even in the case of 
arbitrary week disorder.~\cite{Molenkamp2,Shytov,Khaetskii,Schwab}.
In this communication we provide a fully microscopic calculation of the spin-Hall 
conductivity for the generalized model of non-parabolic spectrum and arbitrary 
momentum dependence of the Rashba velocity~(\ref{oneHam}), within the semiclassical 
theory of disordered conductors, i.e. in the limit $\hbar/\tau\ll \epsilon_F$, 
$\Delta \ll \epsilon_F$. We show, that $\sigma_{sH} = 0$ {\it independently} upon 
the parameter $\tau\Delta/\hbar$ similarly to the case of the parabolic spectrum 
Ref.~\onlinecite{Molenkamp2,Shytov,Khaetskii,Schwab}.

2D isotropic Rashba gas is an electron system with the broken inversion symmetry. 
In this case an electric field perpendicular to the plane could arise. It has no 
effect on the electron orbital motion but it couples to the electron spin via a 
relativistic spin-orbit interaction known as the Rashba term. The Hamiltonian of 
an electron consists of the kinetic energy term and the Rashba term:
\begin{equation}\label{oneHam}
\hat{h}_{\alpha\beta}(\vec{p})=\epsilon(p)\delta_{\alpha\beta}+\alpha(p) \left(
\sigma^x_{\alpha\beta} \hat{p}_{y}-\sigma^y_{\alpha\beta}\hat{p}_{x}
\right)  ,
\end{equation}
where $\alpha(p)$ is the Rashba velocity, $\sigma^i$ ($i=x,y,z$) are the Pauli 
matrices and $\alpha,\beta$ are the spin indices. 
The Hamiltonian~(\ref{oneHam}) can be diagonalized by the unitary matrix:
\begin{equation}\label{Unitary}
U(\vec{p})={1\over\sqrt{2}} \left( \begin{array}{cc} \displaystyle 1 & 1 
\\ \displaystyle ie^{i\varphi_{\bf p}} & -ie^{i\varphi_{\bf p}} \end{array} \right),
\end{equation}
where $\varphi_{\bf p}$ is the angle between the momentum $\vec{p}$ of the electron 
and the $x$-axis, with the eigenvalues:
\begin{equation}\label{eigenv}
\epsilon_\lambda(p)=\epsilon(p) -\lambda p\alpha(p).
\end{equation}
The eigenvalues $\lambda=\pm 1$ of the chirality operator and $\vec{p}$ constitute 
the quantum numbers of an electron state $(\vec{p},\lambda)$. Fermi circles of the 
Rashba gas with the different chiralities are split: 
$p_{F\pm}=p_F(1\pm\alpha(p_F)/v(p_F))$, where Fermi momentum 
$p_F$ solves the equation: $\epsilon(p_F)=\mu$, where $\mu$ is 
the chemical potential; $v(p)=d\epsilon(p)/dp$ 
is the band velocity of the electron. 
The electron velocity in the chiral state is 
$\partial \epsilon_{\lambda}(p)/ \partial p$. 
We assume $\alpha(p_F)\ll v(p_F)$, and neglect corrections of the order
$\alpha/v$. The spin-orbital splitting is then: $\Delta=2p_F\alpha(p_F)$. 
The density of states on the two Fermi circles differs as: 
$\nu_{\pm}= \nu_F(1\pm \alpha_F/v_F)$, where $\nu_F=p_F/2\pi v(p_F)$. 
Contrary to the case of the parabolic spectrum and Rashba velocity independent 
on the momentum, for the generalized model~(\ref{oneHam}) the Fermi velocities 
are different on the two Fermi circles: 
$v_{F+}-v_{F-}=2\alpha_F\left(\frac{p_F}{mv_F}-1\right)-
2p_F\frac{d\alpha}{dp}\Big|_F$.
In the following we use the units where $\hbar=1$. 

We consider the 2D ideal (non-interacting electrons) Rashba electron gas with 
the Hamiltonian:
\begin{equation}\label{mainHam}
\hat{H}_R= \sum_{\vec{p}} a^{\dagger}_{\alpha}(\vec{p})\ 
\hat{h}_{\alpha\beta}(\vec{p})\ a_{\beta}(\vec{p}),
\end{equation}
at zero temperature. $a^{\dagger}_{\alpha}(\vec{p})$ and $a_{\beta}(\vec{p})$ 
are the electron creation and annihilation operators. Electromagnetic vector 
potential $\vec{A}$ couples to the orbital motion of the electron according to 
the transformation: $\vec{p}\rightarrow \vec{p}- e\vec{A}/c$, in the 
Hamiltonian~(\ref{oneHam}). Variation of the Hamiltonian~(\ref{mainHam}) 
with respect to $\vec{A}$ gives the electric current operator: 
$\hat{J}_{\nu}=\sum_{\vec{p}} a^{\dagger}_{\alpha}(\vec{p}) 
(\hat{j}_{\nu})_{\alpha\beta} a_{\beta}(\vec{p})$, where the 
one-particle current operator reads (it is actually a velocity 
$\hat{j}_\nu=e\hat{v}_\nu$):
\begin{equation}\label{Jx}
(\hat{j}_{\nu})_{\alpha\beta}(\vec{p})= 
e\left( v(p)\frac{p_{\nu}}{p}\delta_{\alpha\beta}+ 
\frac{d(p_\mu\alpha(p))}{dp_\nu} \epsilon^{zi\mu}\sigma^i_{\alpha\beta} \right),
\end{equation}
with $\nu=x,y$ being the spatial index and $\epsilon^{zi\mu}$ is the 3D 
totally antisymmetric tensor.

Under the non-uniform SU(2) electron spinor transformation: 
$a_\alpha(\vec{r})\mapsto U_{\alpha\beta}(\vec{r}) a_\beta(\vec{r})$, 
the Hamiltonian~(\ref{mainHam}) becomes dependent on the SU(2) 
``spin electromagnetic'' vector potential 
$\hat{A}_\mu=A^0_\mu\sigma^0+A^i_\mu\sigma^i$, 
where $A^0_\mu$ coincides with the physical electromagnetic potential and 
$A^i_\mu= -i\,\textrm{Tr} (\sigma^iU^+\partial_\mu  U)/2$. 
Although this latter potential is a pure gauge and has no physical consequences, 
variation of the Hamiltonian~(\ref{mainHam}) with respect to it defines 
the spin current of $i$-component of the spin along the direction $\mu$: 
$\hat{J}_{\mu}^i= \sum_{\vec{p}} a^{\dagger}_{\alpha}(\vec{p}) 
(\hat{j}_{\mu}^i)_{\alpha\beta} a_{\beta}(\vec{p})$, where the one-particle 
spin current operator reads:
\begin{equation}\label{Jyz}
(\hat{j}_{\mu}^i)_{\alpha\beta}(\vec{p})= v(p)\frac{p_{\mu}}{p} 
\sigma^i_{\alpha\beta}+ \frac{d(p_\nu\alpha(p))}{dp_\mu} \epsilon^{zi\nu} 
\delta_{\alpha\beta}.
\end{equation}
Our definition of the spin current~(\ref{Jyz}) coincides with the definition 
followed in Ref.~\onlinecite{Rashba}: 
$\hat{J}_{\mu}^i= (\hat{v}_{\mu}\sigma^i+\sigma^i\hat{v}_{\mu})/2$, 
but differs from the definition followed in 
Refs.~\cite{Murakami,Sinova,Molenkamp2} by a factor 2, 
which makes our value of the spin-Hall conductivity being twice as big as that 
in the literature.~\cite{Murakami,Sinova,Molenkamp2}

The interaction of electron with short-ranged non-magnetic 
impurities at positions $\vec{R}_i$, numerated by the index $i$, 
is described by the impurity Hamiltonian:
\begin{eqnarray}\label{Himp}
\hat{H}_{imp}=\sum_i\int u(\vec{r}-\vec{R}_i) a_\alpha^{\dagger}(\vec{r}) 
a_\alpha(\vec{r})d^2\vec{r},
\end{eqnarray}
where $u(\vec{r})$ is a short-range impurity potential. We assume it to be 
sufficiently weak in order for the Born approximation to be valid. 
In this limit the impurity model (\ref{Himp}) is equivalent to the model of 
the Gaussian random potential. We expand the electron Green function 
averaged over the realizations of the disorder potential purturbatively 
in power of the Hamiltonian (\ref{Himp}) using the diagrammatic procedure 
\cite{AGD}. It is a sum of diagrams where a chain of electron bare Green 
functions is separated by impurity ``crosses''. Two ``crosses'' are connected 
by the averaged impurity line: $n_{imp}u^2=1/2\pi\nu\tau$, where $n_{imp}$ is 
the density of impurities, and $\tau$ is the scattering mean free time. 
A ``cross'' does not change the electron spin and the electron frequency 
since the electron scattering off impurity is elastic. Therefore the impurity 
line carries zero frequency. Diagrams with crossings of two or more impurity 
lines are small as powers of the ratio $1/\epsilon_F \tau\ll 1$. 
The averaged Green function is a two by two matrix in the spin space and it is 
a solution of the Dyson equation:
\begin{equation}\label{Dyson}
G_{\alpha\beta}^{-1}(\epsilon,\vec{p})- 
(\epsilon-\mu)\delta_{\alpha\beta}+ h_{\alpha\beta}(\vec{p})= 
-\frac{n_{imp}u^2}{V} \sum_{{\vec{p'}}} G_{\alpha\beta}(\epsilon,\vec{p'}).
\end{equation}
It can be conveniently transformed into the chiral basis by the unitary matrix 
$U(\vec{p})$~(\ref{Unitary}). The retarded and the advanced 
averaged Green functions are diagonal in the chiral basis: 
$G^{(R,A)}_{\lambda'\lambda}(\epsilon,\vec{p})= 
G^{(R,A)}_{\lambda}(\epsilon,\vec{p}) \delta_{\lambda'\lambda}$, 
and the solution to the Dyson Eq.~(\ref{Dyson}) reads~\cite{AGD}:
\begin{equation}\label{Greenfunction}
G^{R}_{\lambda}(\epsilon,\vec{p})= 
\frac{1}{\epsilon-\epsilon_\lambda(\vec{p}) +\mu + i/2\tau}
\delta_{\lambda'\lambda},
\end{equation}
where $\tau$ is explicitely independent of the chirality. 
The advanced Green function is a complex conjugate of the retarded one: 
$G^A_{\lambda}(\epsilon,\vec{p})= \{G^{R}_{\lambda}(\epsilon,\vec{p})\}^*$. 
The Green function in the spin basis is a non-diagonal two by two matrix: 
\begin{equation}\label{GRA}
G^{(R,A)}(\epsilon,\vec{p})=\left(
\begin{array}{cc} G^{(R,A)}_{\uparrow\uparrow}(\epsilon,\vec{p}) & 
G^{(R,A)}_{\uparrow\downarrow}(\epsilon,\vec{p}) \\ \\ 
G^{(R,A)}_{\downarrow\uparrow}(\epsilon,\vec{p}) & 
G^{(R,A)}_{\downarrow\downarrow}(\epsilon,\vec{p}) \end{array} 
\right),
\end{equation}
where (omitting for a moment the frequency and momentum notations):
\begin{eqnarray}
&&G^{(R,A)}_{\uparrow\uparrow}= G^{(R,A)}_{\downarrow\downarrow}= 
(G^{(R,A)}_+ + G^{(R,A)}_- )/2,\nonumber\\
&&G^{(R,A)}_{\uparrow\downarrow}= 
-ie^{-i\varphi_{\bf p}} (G^{(R,A)}_+ - G^{(R,A)}_- )/2, \nonumber\\
&&G^{(R,A)}_{\downarrow\uparrow}= 
ie^{i\varphi_{\bf p}} (G^{(R,A)}_+ - G^{(R,A)}_- )/2,
\end{eqnarray}
with the chiral $G^{(R,A)}_{\pm}$ being defined in 
Eq.~(\ref{Greenfunction}). 

In order to calculate the current, induced in the electron system by 
an electric field, we use the Keldysh technique~\cite{Keldysh}. 
Our result given by the Eq.~(\ref{Nresponse}) is well known but we 
derive it here for consistency. The averaged Keldysh Green function 
is a four by four matrix $\mathcal{G}(p,\epsilon)$ that can be conveniently 
factorized into a two by two Keldysh matrix whose elements are matrices 
in the spin space themselves:
\begin{eqnarray}\label{G_Keldysh}
&&\left(
\begin{array}{cc} \mathcal{G}_{--} & \mathcal{G}_{-+} \\ \mathcal{G}_{+-} 
& \mathcal{G}_{++} \end{array} 
\right)= \left(
\begin{array}{cc} 1-N(\epsilon) & -N(\epsilon)\\ 1-N(\epsilon) 
& -N(\epsilon) \end{array} 
\right) G^R(p,\epsilon)+ \nonumber\\ 
\nonumber \\
&&+\left(\begin{array}{cc} N(\epsilon) & N(\epsilon)\\ -1+N(\epsilon)
&-1+N(\epsilon) \end{array} 
\right)G^A(p,\epsilon), 
\end{eqnarray}
where the electron distribution $N(\epsilon)$ is proportional to the 
unit matrix in the spin space.

\begin{figure}
\includegraphics[angle=0,width=0.26\textwidth]{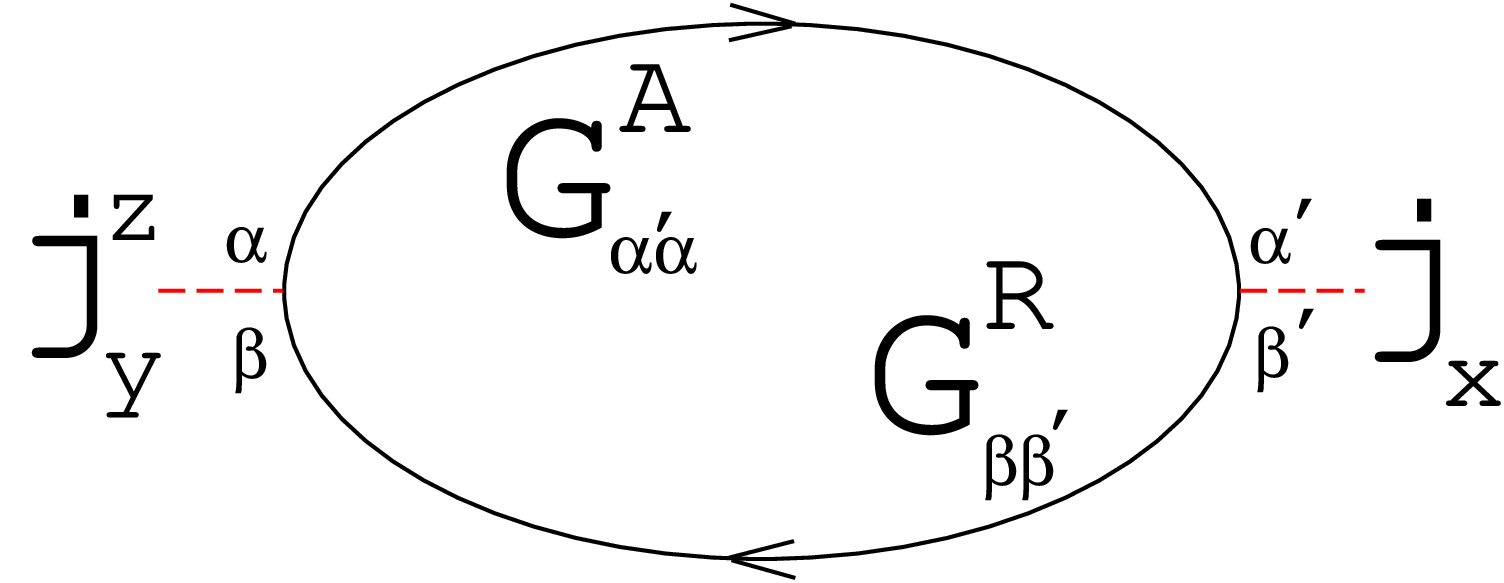}
\caption{\label{fish} The spin-Hall conductivity 
is given by a one-loop diagram.}
\end{figure}

\begin{figure}
\includegraphics[angle=0,width=0.3\textwidth]{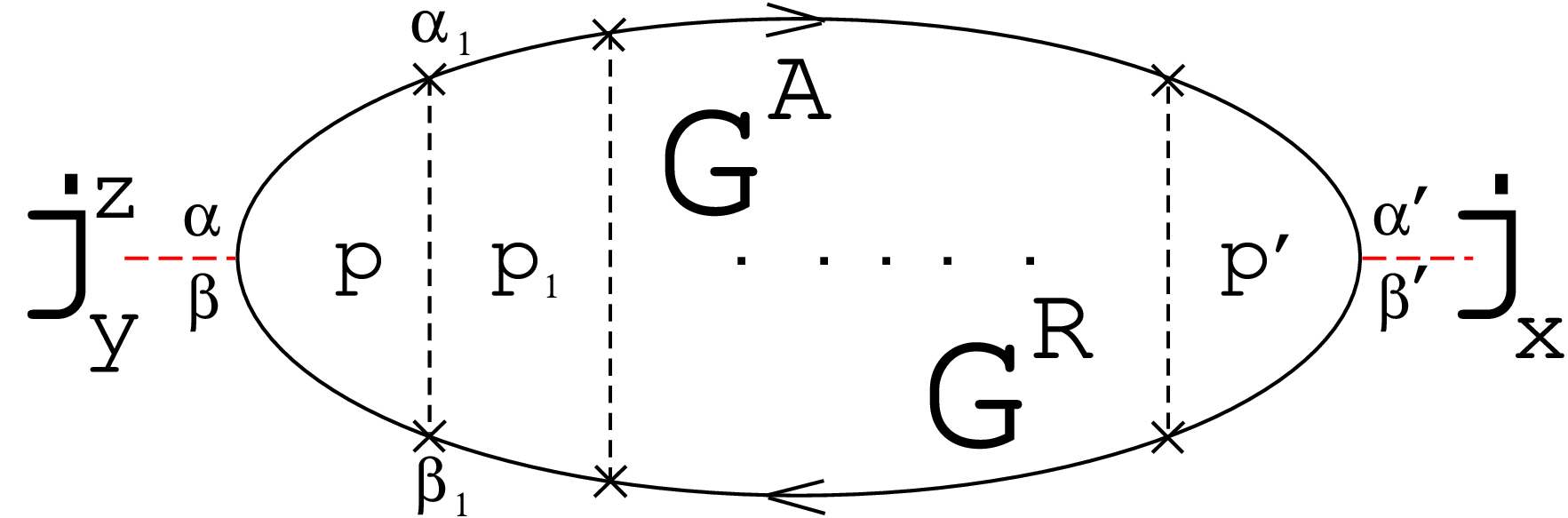}
\caption{\label{ladder} The vertex correction to the spin-Hall conductivity is 
given by a sum of non-crossing ladder diagrams.}
\end{figure}

We choose the gauge for the uniform electric field 
$\vec{E}(t)=\vec{E}(\Omega) e^{-i\Omega t}$ to be a time dependent 
vector potential $\vec{A}(t)=\vec{A}(\Omega) e^{-i\Omega t}$, where 
$\vec{A}(\Omega)= -ic\vec{E}(\Omega)/\Omega$. 
Using the Keldysh technique~\cite{Keldysh} we average the spin current 
operator over the electron state perturbed by the electromagnetic 
Hamiltonian: $\hat{H}_{em}=
-\frac{1}{c} \int d^2\vec{r} \hat{j}_{\nu}(\vec{r}) A_{\nu}(t)$, 
in the first order of the perturbation theory. The spin-Hall 
conductivity $\sigma_{sH}$ is then found from the relationship: 
$\langle\hat{j}_{\mu}^z(\Omega)\rangle= 
\epsilon_{\mu\nu} \sigma_{sH}(\Omega) E_{\nu}(\Omega)$, as:
\begin{equation}\label{response_Keldysh}
\sigma_{sH}=\frac{-1}{V\Omega} \sum_{\vec{p}} 
\int \frac{d\epsilon}{2\pi} \textrm{Tr} \left[\hat{j}_y^z(\vec{p}) 
\mathcal{G}(\epsilon+\Omega,\vec{p})\tau^z \hat{j}_x(\vec{p}) 
\mathcal{G}(\epsilon,\vec{p}) \right]_{-+}
\end{equation}
where $\tau^z$ is the four by four matrix given by the direct product 
of the Pauli matrix $\sigma^z$ in the Keldysh space and the unit matrix 
in the spin space, the current operators in Eq.~(\ref{response_Keldysh}) 
are the direct product of matrices~(\ref{Jx},~\ref{Jyz}) and the unitary 
matrix in the Keldysh space. $\textrm{Tr}$ in Eq.~(\ref{response_Keldysh}) 
operates only in the spin space whereas the indices of $-+$ element 
corresponds to the Keldysh space. Substituting the Green function 
Eq.~(\ref{G_Keldysh}), we obtain:
\begin{eqnarray}\label{Nresponse}
\sigma_{sH}(\Omega)=
\frac{1}{V\Omega}\sum_{\vec{p}}\int\frac{d\epsilon}{2\pi}\Big\langle 
\textrm{Tr} 
\Big[ \hat{j}_y^z(\vec{p})\Big(N(\epsilon+\Omega) \nonumber\\  
{}\times \left(G^R(\epsilon+\Omega,\vec{p})-
G^A(\epsilon+\Omega,\vec{p})\right) 
\hat{j}_x(\vec{p})G^A(\epsilon,\vec{p})  \nonumber\\ 
{}+G^R(\epsilon+\Omega,\vec{p})\hat{j}_x(\vec{p}) N(\epsilon)
\left(G^R(\epsilon,\vec{p})-G^A(\epsilon,\vec{p})\right) \Big)\Big] 
\Big\rangle,
\end{eqnarray}
where the brackets indicate averaging over the disorder. 
In the non-crossing approximation the average in Eq.~(\ref{Nresponse}) is 
given by the sum of the one-loop and the ladder diagrams shown in 
Figs.~\ref{fish},~\ref{ladder}.

First, we calculate the one-loop diagram Fig.~\ref{fish} and denote this 
part of the spin-Hall conductivity as $\sigma_{sH}^0$. It corresponds to 
the Eq.~(\ref{Nresponse}) with all Green functions being substituted by 
the averaged Green functions~(\ref{Greenfunction}). The second line in 
Eq.~(\ref{Nresponse}) contains the imaginary part of the Green function 
$G^R(\epsilon+\Omega,\vec{p})$ and the corresponding integral is convergent, 
therefore we change $\epsilon$ to $\epsilon-\Omega$ in this second line. 
At $T=0$ the Fermi-Dirac distribution function reads: 
$N(\epsilon)= \theta(-\epsilon)$. We take the integral over 
$\epsilon$ and in the zero-frequency limit $\Omega\rightarrow 0$ we find:
\begin{eqnarray}\label{spH}
\sigma^0_{sH}=-\frac{1}{V}\sum_{\vec{p}} \frac{v(p)}{2\pi p}
\Big[\frac{1}{2p\alpha(p)}\left(S(\zeta)-S(\xi)\right)+ \nonumber\\
 +\frac{1}{2\tau}\frac{\zeta\xi+1/4\tau^2}{(\zeta^2+1/4\tau^2)
 (\xi^2+1/4\tau^2)}\Big],
\end{eqnarray}
where $S(x)=\arctan(x/2\tau)$, $\zeta=\epsilon(p)-p\alpha(p)-\mu$ and 
$\xi=\epsilon(p)+ p\alpha(p)-\mu$. In the large volume limit we substitute 
$\frac{1}{V}\sum_{\vec{p}}\rightarrow \int \frac{d^2\vec{p}}{(2\pi)^2}$, 
and then evaluate the integral over $\vec{p}$ in~(\ref{spH}) in the limit 
of large Fermi circle $\mu\gg \Delta, 1/\tau$. This procedure, known as the 
semiclassical approximation, expresses the momentum $p$ in terms of the 
quasiparticle energy $\xi$:
\begin{equation}\label{semi}
\int\frac{d^2\vec{p}}{(2\pi)^2}\approx \frac{p}{2\pi v(p)}
\int_{-\infty}^{\infty} \left(1+R(p) \xi\right) d\xi 
\int_{0}^{2\pi}\frac{d\varphi}{2\pi},
\end{equation}
where $R(p)=p^{-1}v^{-1}(p)-m^{-1}(p)v^{-2}(p)$. The result reads:
\begin{equation}\label{loopResult}
\sigma^{0}_{sH}=\frac{e}{4\pi}\left(1-\frac{1}{1+(\Delta\tau)^2}\right),
\end{equation}
with the first and the second terms corresponding to the first and the second 
terms of Eq.~(\ref{spH}), respectively.

An important observation is that the result~(\ref{loopResult}) coincides 
exactly with the result obtained from those terms in Eq.~(\ref{Nresponse}) 
that contains one retarded and one advanced Green functions. These two terms 
are proportional to $dN(\epsilon)/d\epsilon= -\delta(\epsilon)$ 
and all integrals 
are explicitely confined to the vicinity of the Fermi circle. 
Therefore the 
spin-Hall conductivity unlike the usual Hall conductivity is 
determined by 
the quasi-particles around the Fermi circle and not by the 
entire Fermi disk.

The ladder diagrams shown in Fig.~\ref{ladder} represent the 
vertex corrections 
to the current. Additional impurity lines improve the 
convergence of the integral 
in Eq.~(\ref{Nresponse}). As a consequence, vertex corrections 
to the terms 
in Eq.~(\ref{Nresponse}) with the two advanced or with the two 
retarded Green 
functions vanish as $\textrm{max} (1/\tau,\Delta)/ \epsilon_F\ll 1$. 
Therefore 
we consider only the vertex corrections to the terms with one 
advanced and one 
retarded Green's functions as it is shown in Fig.~\ref{ladder}. 
For these diagrams the semiclassical approximation~(\ref{semi}) is valid.

\begin{figure}
\centering
\includegraphics[angle=0,width=0.33\textwidth]{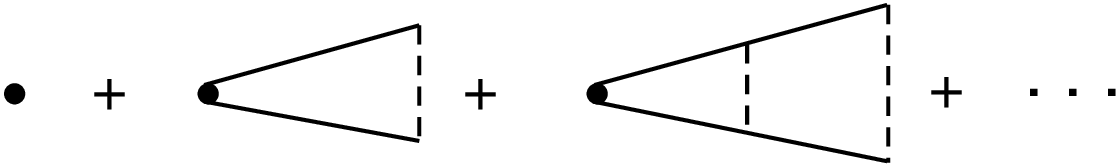}
\caption{\label{vertex} The vertex of the spin current with the 
vertex corrections taken into account: $j_y^z(\vec{p})+\tilde{J}_y^z$.}
\end{figure}

The sum of ladder diagrams with $n=1, 2,...$ impurity lines is given by the 
expression:
\begin{equation}\label{response_ladder}
\sigma_{sH}^{\text{lad}}=-\int \frac{d^2\vec{p}}{(2\pi)^2}\textrm{Tr} 
\left[\tilde{J}_y^z G^R(0,\vec{p}) j_x(\vec{p}) G^A(0,\vec{p})\right],
\end{equation}
where the sum of $n=1..\infty$ vertex corrections to the current 
$\hat{j}_y^z(\vec{p})$  (with at least one impurity line) is denoted by the 
matrix $\tilde{J}_y^z$. In the spin basis and for short-ranged impurity 
potentials it does not depend on the electron momentum $\vec{p}$ and satisfies 
the transfer matrix equation:
\begin{equation}\label{vertex_cor}
\tilde{J}_y^z=\frac{1}{2\pi\tau\nu}\int\frac{d^2\vec{p}}{(2\pi)^2} 
G^A(0,\vec{p})\left[j_y^z(\vec{p})+ \tilde{J}_y^z\right]G^R(0,\vec{p}),
\end{equation}
where the Green functions $G^{(R,A)}$ are given by Eq.~(\ref{GRA}). 
The ``full'' spin current operator with all vertex corrections included: 
$\hat{j}_y^z(\vec{p})+\tilde{J}_y^z$, is represented diagrammatically 
in Fig.~\ref{vertex}. In the equations for the current 
operators~(\ref{Jx},~\ref{Jyz}) we expand the electron velocity 
$v(p)=v(p_F)+ \xi/(v(p_F)m(p_F))$, where $m^{-1}(p)=dv(p)/dp$, 
to the first order in the deviation from the Fermi circle: $\xi/\mu$, 
small in the semiclassical approximation. We also expand the spin-orbital 
splitting: 
$p\alpha(p)=
\alpha(p_F)\left(p_F+(1+(p_F/\alpha_F)d\alpha/dp_F)\xi/v_F\right)$, 
in the Green functions. We then evaluate Eq.~(\ref{vertex_cor}) 
in the semiclassical approximation~(\ref{semi}) 
neglecting odd powers of $\xi$:
\begin{eqnarray}\label{vc}
&&\left(\tilde{J}_y^z\right)_{\uparrow\uparrow}= 
\left\{ \left[2+(\Delta\tau)^2\right] 
\left(\tilde{J}_y^z\right)_{\uparrow\uparrow} +
(\Delta\tau)^2\left(\tilde{J}_y^z\right)_{\downarrow\downarrow} \right\}B,
\nonumber\\ &&\left(\tilde{J}_y^z\right)_{\downarrow\downarrow}=
\left\{ (\Delta\tau)^2 \left(\tilde{J}_y^z\right)_{\uparrow\uparrow}+ 
\left[2+(\Delta\tau)^2\right] 
\left(\tilde{J}_y^z\right)_{\downarrow\downarrow} \right\}B,
\nonumber\\ &&\left(\tilde{J}_y^z\right)_{\uparrow\downarrow}=
\left\{ -iv(p_F)\Delta\tau+ \left[2+(\Delta\tau)^2\right] 
\left(\tilde{J}_y^z\right)_{\uparrow\downarrow} \right\}B,
\nonumber\\ &&\left(\tilde{J}_y^z\right)_{\downarrow\uparrow}=
\left\{ iv(p_F)\Delta\tau+ \left[2+(\Delta\tau)^2\right] 
\left(\tilde{J}_y^z\right)_{\downarrow\uparrow} \right\}B,
\end{eqnarray}
where $B=\frac12 \left[1+(\Delta \tau)^2\right]^{-1}$. 
From the first two lines of Eqs.~(\ref{vc}) we find: 
$(\tilde{J}_y^z)_{\uparrow\uparrow}= 
(\tilde{J}_y^z)_{\downarrow\downarrow}$, 
whereas from the last two lines we find: 
$(\tilde{J}_y^z)_{\downarrow\uparrow}=iv(p_F)/(\Delta\tau)$ and 
$(\tilde{J}_y^z)_{\uparrow\downarrow}= -iv(p_F)/(\Delta\tau)$. 
The integrand of Eq.~(\ref{response_ladder}) does not depend on 
the diagonal elements of the matrix $\tilde{J}_y^z$ and therefore we set 
them to zero: $\tilde{J}_y^z= \sigma^y\ v(p_F)/\Delta\tau$. 
As it was expected the vertex corrections are proportional 
to the scattering rate. Integrating Eq.~(\ref{response_ladder}) 
in the semiclassical approximation~(\ref{semi}) over $\xi$, we finally 
find the ladder part of the spin Hall conductivity (Fig.~\ref{ladder}):
\begin{equation}\label{spinHallladder}
\sigma_{sH}^{\text{lad}}= 
\frac{e}{4\pi}\left(-1+\frac{1}{1+(\Delta\tau)^2}\right).
\end{equation}
Remarkably all derivatives of $\alpha(p)$ and $v(p)$ over 
$p$ have canceled out.

The spin-Hall conductivity is the sum of 
Eqs.~(\ref{loopResult},~\ref{spinHallladder}) and is zero:
\begin{equation}\label{SHC}
\sigma_{sH}=\sigma_{sH}^{0}+\sigma_{sH}^{\text{lad}}= 0.
\end{equation}
It explicitely does not depend on the impurity scattering time $\tau$. 
But, we observe a discontinuity between the spin-Hall 
conductivity in the clean system $\sigma_{sH}= e/8\pi\hbar$ and the 
spin-Hall conductivity Eq.~(\ref{SHC}) in the presence of the 
infinitely small amount of non-magnetic scatterers. 
As it was shown in Ref.~\onlinecite{Schwab}, this discontinuity 
is related to the dissipation in the system, which gives rise to the 
dissipative part in the spin-Hall conductivity $\sigma_{sH}^D=-e/8\pi\hbar$,
which cancels the reactive part $\sigma_{sH}^R=e/8\pi\hbar$.

In an analogous calculation, for the generalized model~(\ref{oneHam}), 
we find the magneto-electric effect - 
magnetization induced by the electric field:
\begin{equation}\label{magnetization}
\langle \hat{S}^\mu\rangle=\epsilon^{\mu\nu}\frac{e\Delta\tau}{2\pi v_F}E_\nu,
\end{equation}
where 
$\hat{S}^\mu=1/2\ \sum_{\vec{p}} 
a^{\dagger}_{\alpha}(\vec{p})\ 
\hat{\sigma}^\mu_{\alpha\beta}a_{\beta}(\vec{p})$ is the total spin operator
and the total magnetization $\langle \hat{S}^y\rangle$ is in agreement 
with Ref.~\onlinecite{Shytov,Edelstein,Aronov}. 
One should notice that the steady in-plane magnetization~(\ref{magnetization}) 
is a consequence of the zero spin-Hall effect~(\ref{SHC}).

Non-zero spin Hall conductivity would result in a non-steady in-plane 
magnetization. This follows from the evolution equation of the total 
spin of the system $\hat{S}_{\nu}$ and the commutation relationship:
\begin{equation}\label{totalspin}
-i\frac{d}{dt}\hat{S}^\mu=[\hat{H},\hat{S}^{\mu}]= 
i\alpha(p)\frac{p}{v(p)}\hat{J}_{\mu}^z,
\end{equation}  
where $\mu=x,y$; $\hat{J}_{\mu}^z$ is the total spin current operator, 
defined in~(\ref{Jyz}). Remarkably, Eq.~(\ref{totalspin}) 
is valid for an electron system 
with any non-magnetic disorder and any electron-electron interaction. 
Moreover, valid is a more general equation for the local spin density evolution:
\begin{equation}\label{localspin}
\frac{\partial}{\partial t} \hat{s}^\mu(\vec{r})+
\frac{1}{2}\frac{\partial}{\partial x_\nu} \hat{j}^\mu_\nu+
\alpha m\left(\epsilon^{\mu yl}\hat{j}_x^l-
\epsilon^{\mu xk}\hat{j}_y^k\right)=0,
\end{equation}
where $\mu=x,y,z$; in derivation of Eq.~(\ref{localspin}) we considered 
the standard model of parabolic spectrum.
If the last term were zero then the above Eq.~(\ref{localspin}) 
would be the spin conservation equation. But in the Rashba metal 
neither component of the spin is conserved.

To conclude, we have extended the result of Inoue et al. 
Ref.~\onlinecite{Molenkamp2} $\sigma_{sH}=0$ for the case of the 
arbitrary electron dispersion, arbitrary strength of disorder and 
arbitrary momentum dependence of the Rashba velocity $\alpha(p)$. 
Our result agrees with Ref.~\onlinecite{Shytov,Khaetskii,Schwab}. 

The author thanks M. V. Feigel'man and E. I. Rashba for illuminating discussions.  
I am grateful to A. V. Shytov for pointing out a mistake in my previous 
calculation. This research was supported by Dynasty Foundation,  
Landau Scholarship-Juelich, 
Program "Quantum Macrophysics" of Russian Academy of Sciences, 
and the RFBR grant \# 04-02-16348.

\end{document}